\documentclass{osa-article}

\journal{osajournal}

\articletype{Research Article}

\begin{document}

\title{Higher order harmonic generation and strong field ionization with Bessel-Gauss beams in a thin jet geometry}

\author{Michael Davino,\authormark{1} Adam Summers,\authormark{2,3} Tobias Saule,\authormark{1} Jan Tross,\authormark{2} Edward McManus,\authormark{1} Brandin Davis,\authormark{1} and Carlos Trallero-Herrero\authormark{1,2,*}}

\address{\authormark{1}Department of Physics, The University of Connecticut, 196 Auditorium Road, Storrs, CT 06269, USA}
\address{\authormark{2}J. R. Macdonald Lab., Kansas State University, 116 Cardwell Hall, Manhattan, KS 66506, USA}
\address{\authormark{3}CFO, Institut de Ciencies Fotoniques, Mediterranean Technology Park, Av. Carl Friedrich Gauss 3, 08860, Castelldefels, Barcelona, Spain}

\email{\authormark{*}carlos.trallero@uconn.edu} 



\begin{abstract}
A promising alternative to Gaussian beams for use in strong field science is Bessel-Gauss (BG or Bessel-like) laser beams as they are easily produced with readily available optics and provide more flexibility of  the spot size and working distances. Here we use BG beams produced with a lens-axicon optical system for higher order harmonic generation (HHG) in a thin gas jet. The finite size of the interaction region allows for scans of the HHG yield along the propagation axis. Further, by measuring the ionization yield in unison with the extreme ultraviolet (XUV) we are able to distinguish regions of maximum ionization from regions of optimum XUV generation. This distinction is of great importance for BG fields as the generation of BG beams with axicons often leads to  oscillations of the on-axis intensity,which can be exploited for extended phase matching conditions. We observed such oscillations in the ionization and XUV flux along the propagation axis for the first time. As it is the case for Gaussian modes, the harmonic yield is not maximum at the point of highest ionization. Finally, despite Bessel beams having a hole in the center in the far field, the XUV beam is well collimated making BG modes a great alternative when spatial filtering of the fundamental is desired.
\end{abstract}

\section{Introduction}

With the discovery of high harmonic generation (HHG) in the late 1980s and 1990s \cite{McPherson1987,LHuillier1993, Krause1992,Corkum1993}, and the proceeding evolution of HHG based light sources \cite{Corkum2007,Krausz2009a}, a variety of new fields of physics were developed. HHG has allowed for the ability to produce pulses of coherent radiation with pulse durations of less than 100 attoseconds \cite{Gaumnitz2017} and with photon energies exceeding 1keV \cite{Popmintchev2010}. These light sources unlocked the ability to probe ultrafast processes, such as photoionization \cite{Goulielmakis2010,Uiberacker2007,Schultze2010} chemical bond rearrangement \cite{Attar2017,Geneaux2019,Kraus2018} and electronic dynamics in solid-state materials \cite{Geneaux2019,Cavalieri2007}, on their natural time scales.  Additionally, the multi-octave spectral bandwidth inherently present in XUV and X-ray attosecond pulses has given the ability to probe matter with elemental specificity across a wide range of absorption edges \cite{Silva2015}. As such, the development of HHG light sources with higher flux, higher photon energy and shorter pulse duration has received immense and continued interest \cite{Young2018}. 
\\
A multitude of approaches to produce more advanced HHG light sources have been pursued. These have included novel driving laser sources, target geometries, and generation media \cite{Krausz2009a,Corkum2007,Popmintchev2010,Vampa2015}. However, comparatively little work has been undertaken to explore the advantages of using non-Gaussian mode profiles in the driving laser source. 
\\
In this work we  explore the use of non-Gaussian beams (i.e. non-Gauss-Hermite) in HHG. Here we specifically investigate using Bessel-Gauss (BG) beams \cite{McLeod1954,McLeod1960}, also known as Bessel-like \cite{Durnin1987,Durnin1988}, for the generation of the XUV radiation. In the past we have characterized the properties of such beams and their relevance for strong field science \cite{Summers2017}.
\\
BG beams have a series of enticing properties \cite{akturk2012tailored,Mcgloin2005}. Among the most notable is the ability to overcome limitations of the Rayleigh range \cite{Jarutis2000,Grunwald2008}. Bessel-like beams maintain focused, on-axis peak intensity for much longer distances than focused Gaussian beams \cite{Summers2017}. This allows for the use of an extended generation target geometry as well as potentially more favorable phase matching conditions. Finding and establishing the best conditions for phase matching has been one of the most active areas of high harmonic and attosecond science \cite{lin_le_jin_wei_2018}.  The potential phase matching benefits from Bessel-like beams primarily arise from the difference in on-axis phase front velocity as well as the extended intensity profile that can be achieved using BG beams for HHG as opposed to conventional Gaussian focal profiles \cite{Vampa2015}. Further, recent theoretical work on two-color gating with BG beams shows promise for the generation of attosecond pulses of high quality beam profiles \cite{Luo2015,Wang2012,Li2012} or with extended phase matching conditions \cite{Li2012}. 
\\
Experimental studies have previously been presented exploiting the generation of XUV pulses also with a lens-axicon optical system in a semi-infinite gas cell \cite{VanDao2009}. It was demonstrated that BG beams are not only capable of producing XUV pulses but also that they show an increase in XUV flux for some harmonics as compared to standard Gaussian beams which can be attributed to an improvement in phase matching conditions \cite{Averchi2008}.
\\
An additional advantage of Bessel-like beams for HHG is self-filtering. In most experiments, utilizing the radiation produced by HHG requires the elimination of the fundamental beam. This is typically accomplished by employing a series of very thin filters. While these filters block significantly more IR than XUV or X-ray radiation, they still reduce the total flux of the HHG beam. Using transmissive filters for filtering the driving radiation from the HHG radiation also presents a potential problem as HHG pulse durations approach the zeptosecond regime. The limitation with this technique is clear with the realization that the optical path of 100 zs is 30 pm, or roughly two atomic layers of $\alpha$-quartz. Such variation could happen in thin metallic filters due to strain in vacuum or heat because of laser interaction. While some promising work in self-filtering schemes has been presented in using annular beams \cite{Klas2018}, these beams lack the substantial tunability of BG beams produced by a lens-axicon system which includes an extended effective Rayleigh range, focal distance, and spot size \cite{Summers2017}. Further, the focusing of annular beams for HHG results in a decrease in harmonic yield relative to a purely Gaussian beam whereas BG have shown an increase in harmonic yield in a semi-infinite geometry \cite{VanDao2009}.
\\
In this paper we show the first HHG obtained from a capillary (nozzle) where the fundamental is focused using a lens-axicon optical system. The capillary produces a thin gas target region and allows us to explicitly study how HHG flux evolves along the focal profile of the BG beam. This is in contrast to extended gas cell type geometries that do not allow for position sensitive yield measurements. Measuring the local yield for a more complete understanding the interplay between laser intensity and HHG phase matching. This is vital information when working with the complex focal profiles present in Bessel-Gauss beams. The total ionization yield is simultaneously measured with the harmonics, giving the ability to disentangle the focal intensity profile from local phase matching conditions. We have shown in the past that such parallel measurements provide an unambiguous determination of the maximum peak intensity and the relative position of the optical beam and gas jet. Mapping the HHG yield from such beams is an important step in implementing BG beams for high-flux HHG sources.

\section{Experimental methods and results}

\begin{figure}[h!]
\centering\includegraphics[width=105mm]{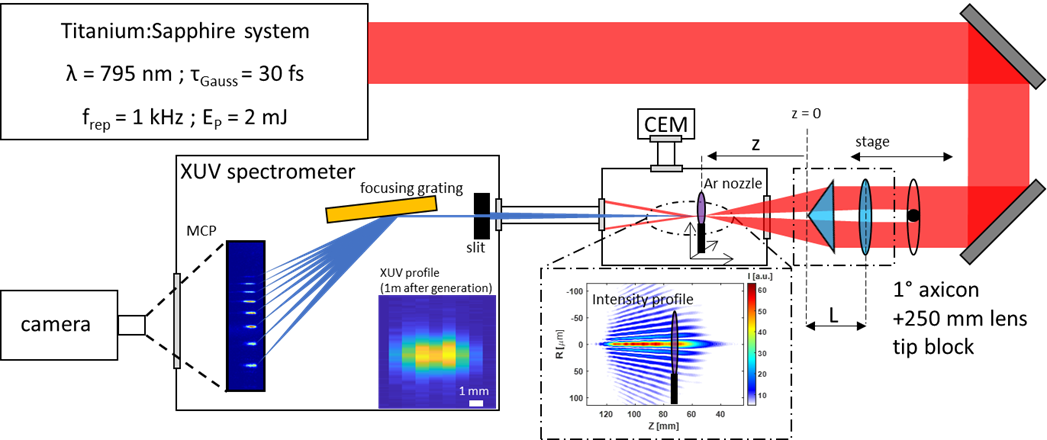}
\caption{Experimental setup. The output of a 1 kHz Ti:Sapph ($\lambda$ = 795 nm, $\tau$ = 30 fs) laser system is incident on a tip block, lens, and axicon producing a BG beam to be focused into an Argon (Ar) gas jet. The XUV beam produced from the gas jet is inherently isolated from the fundamental light as the on-axis BG beam vanishes in the far field leaving only the XUV to enter the spectrometer comprising of a slit, focusing grating, MCP and phosphor screen. The phosphor is imaged by an external camera. To attain a profile of the XUV beam the grating and slit are translated perpendicular to the beam. Also shown is a simulated intensity profile of a femtosecond BG beam.}
\end{figure}

A schematic of the experimental setup is shown in Figure 1. The laser source used for this study is a 1 kHz Titanium:Sapphire system producing 2 mJ pulses with a duration of 30 fs and a central wavelength of 795 nm. Next, to prepare the Bessel-Gauss beam for use in generating XUV, the laser output is sent into an optical setup using a lens and axicon in combination \cite{Summers2017,VanDao2009} with a 250 mm focusing lens set a variable distance L before a one-degree base angle axicon. This versatile geometry allows for studying the effects of different BG focusing conditions on the process of HHG. A 2 mm diameter tip block is placed before the lens and axicon to mitigate the imperfections of the tip of the axicon (as done by S. Akturk et al. \cite{Akturk2008}). Both the lens and axicon are mounted on a computer-controlled stage to allow the focal profile (see inset of Figure 1) to be scanned through the gas jet. The stage has a 30 mm travel range allowing the distance from the axicon tip to the gas jet to be varied between 95 mm and 125 mm. For larger values of L this distance is further limited by the spot size of the beam on the entrance window to the HHG chamber, which induces nonlinear effects as the peak intensity increases. 
\\
Argon gas is injected into the source chamber as a thin gas jet produced by a hollow core (250 µm core diameter) capillary mounted on a 3D manipulator approximately 5 cm from the entrance window of the chamber. The core diameter of the capillary was chosen to be sufficiently smaller than the length scale on which oscillations in on-axis intensity occur in BG beams \cite{Summers2017} (see Figure 1 inset). The Argon backing pressure for this experimental study was approximately 2-3 bar. To keep the gas flow low a needle / leak valve is placed in the gas line, limiting the gas flow to 200 mbar L/sec.
\\
A channel electron multiplier (CEM, channeltron) is placed several centimeters behind the gas jet and slightly offset to allow for simultaneous measurement of ionization yield and HHG signal. This technique was used to demonstrate phase matching conditions in the past \cite{Shiner,Shiner2009}. The source chamber is kept at 6.66 x 10\textsuperscript{-5} mbar during operation by a 2000 L/s magnetic levitation turbo pump to prevent reabsorption in the ambient backing pressure \cite{Henke1993}. \\
An HHG spectrometer (same as in references \cite{tross2017,Trob2019}) is placed approximately 1 m behind the generation region. Differential pumping tubes between the source chamber and the spectrometer allow for two orders of magnitude pressure difference. The spectrometer consists of a slit followed by a curved, variable-spacing, holographic grating (groove density 1,200 grooves/mm), that disperses the harmonics across a Z-stack MCP (Photonis, 105 mm by 35 mm) with a P46 phosphor screen on the output. The slit and grating are mounted on a translational stage allowing a beam profile of the XUV to be taken (see inset of Figure 1). A Hamamatsu Orca Flash 2.8, low-noise, 12-bit resolution camera is used to capture the harmonic spectrum. An example of a raw XUV spectrum is shown in Figure 2. The integration time was 300 ms and harmonics were observed up to the 27th order.
\\
The ionization yield and the XUV spectra were taken simultaneously while varying the nozzle position in a 1D scan along the beam propagation with respect to the lens and axicon. Figure 3 shows two scans for different experimental conditions with L = 13 mm (a) and L = 12.3 mm (b). The top plot is a trace of the ionization yield with the HHG yield attained from integrating the shown 2D map of the harmonic spectrum in Figure 3 (bottom). Varying the distance, L, between the lens and axicon alters the focusing conditions of the BG beam. This can be seen in both the ionization traces as well as the HHG yield dependence as a function of nozzle position.

\begin{figure}[h!]
\centering\includegraphics[width=80mm]{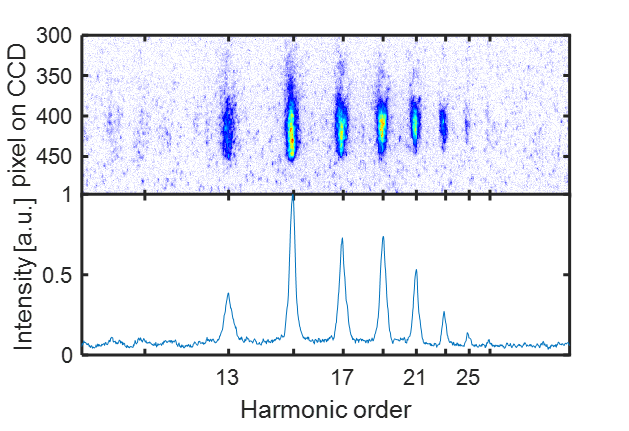}
\caption{Typical spectrum from imaging the MCP/Phosphor screen. The integration time was 300 ms. Harmonics up to the 27th order were observed corresponding to 41 eV.}
\end{figure}

\begin{figure}[h!]
\centering\includegraphics[width=140mm]{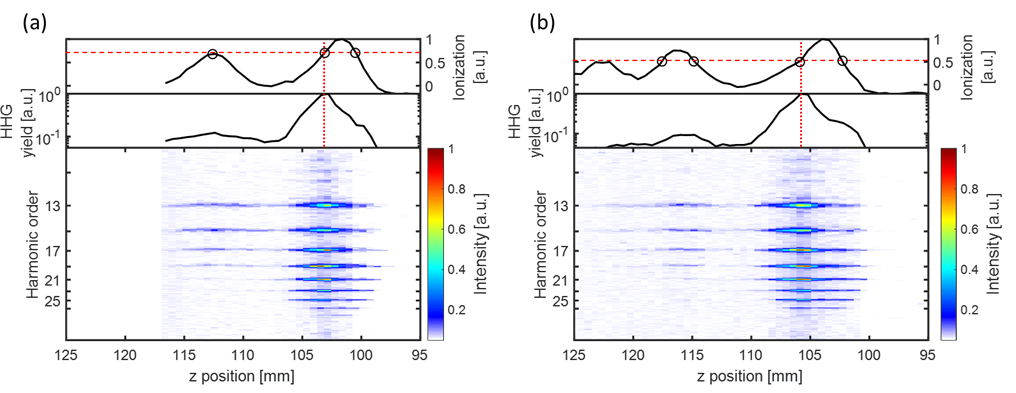}
\caption{From top to bottom; traces of ionization and HHG yield as well as 2D maps of HHG yield for L = 13 mm (a) and L = 12.3 mm (b). Images were taken at 37 (a) and 53 (b) different positions along z, the axis of propagation. At each position the HHG output spectrum was recorded to produce the 2D map. This map was then integrated to produce the trace of the full HHG yield over all detected harmonic orders. Vertical and horizontal red dashed lines in both (a) and (b) indicate the position of maximum HHG yield and the ionization yield at that position respectively. Black circles indicate other positions at which the ionization level was the same as at the position of max XUV flux.}
\end{figure}

Several striking features are immediately visible. The first is the long range over which strong-field ionization takes place (> 2 cm in Figure 2 (b)). The second major feature in the ionization yield is the oscillations as the beam is scanned through the jet. Measurement of the ionization yield gives us direct access to the number of ions produced in the strong field. Much like the intensity profile shown in the Figure 1 inset, oscillations are observed in both the HHG and ionization yields along the axis of the BG beam.  Both the extended ionization range and the oscillations are features absent from strong-field ionization using Gaussian beams.
\\
Oscillations are also present in the HHG data. However, for both the L = 13 mm and L = 12.3 mm cases the HHG yield does not follow with a one-to-one correspondence to the ionization.  In both Figure 3 (a) and Figure 3 (b) we observe that the ionization yield drops to approximately half the peak value in the second maximum while the HHG yield drops by an order of magnitude respectively.  Further, horizontal red dashed lines in Figure 3 (a) and (b) indicate the level of ionization at the focal position of max XUV flux. There are multiple positions at which this same ionization yield is found (marked by black circles), but each with significantly less XUV yield.   These discrepancies in correlation between the harmonic yield and ionization yield point to the presence of different phase matching conditions through different regions, that are not only a function of the laser intensity or ionization rate. Considering the expected on-axis intensity oscillations for BG beams, one might also expect to observe the resulting oscillations in both the XUV and ionization yield. However, the results shown in Figure 3 indicate an added degree of complexity to understanding the full picture for HHG.
\\

\begin{figure}[h!]
\centering\includegraphics[width=130mm]{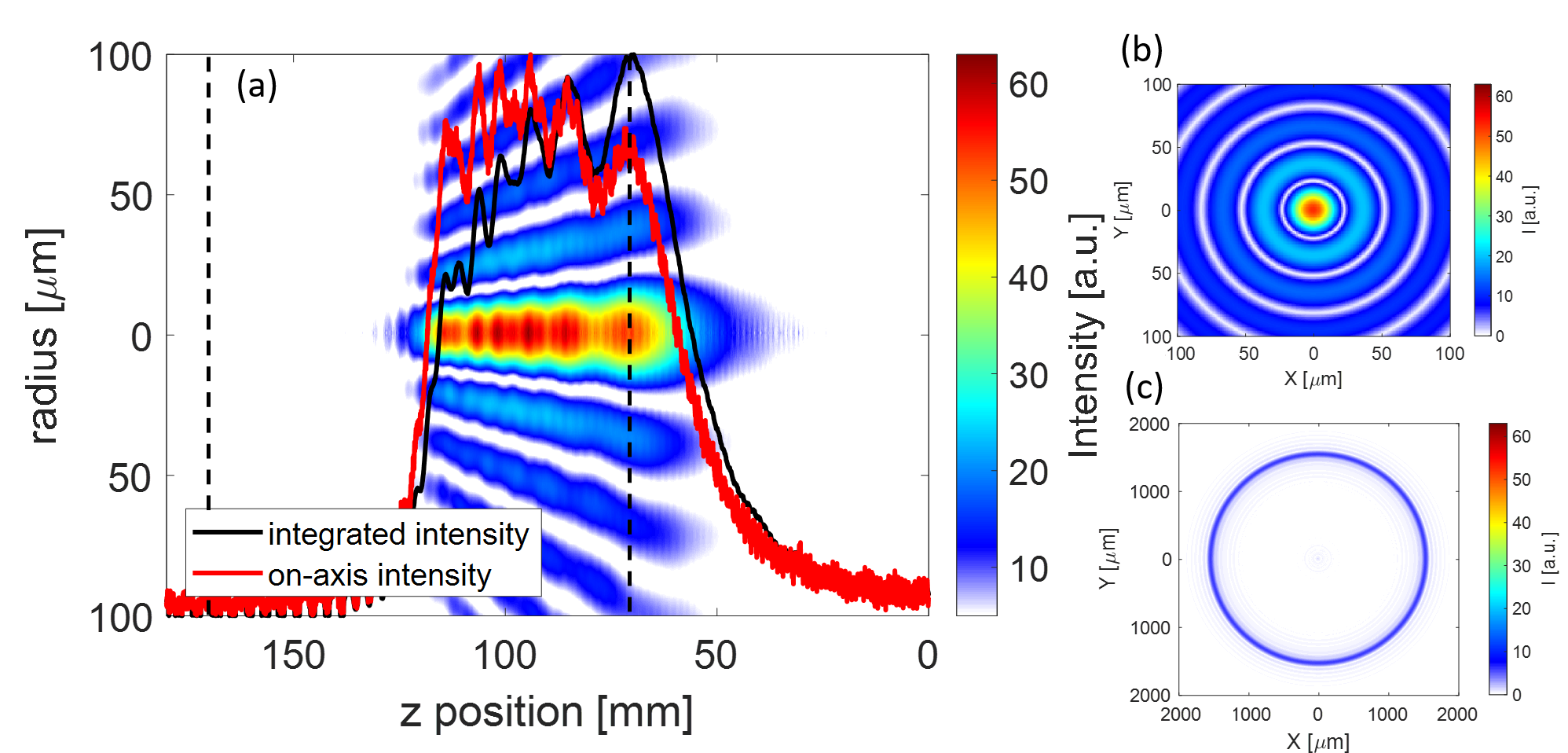}
\caption{Simulated intensity profiles of a BG beam from a lens-axicon system with parameters: 1 mm tip block, L = 12 mm, and a 4 mm FWHM spot size incident on the axicon that closely resembles the experimental conditions. (a) shows intensity as a function of the distance from the tip of the axicon (z, increasing from right to left) and the radial distance. Two line profiles derived from the distribution in (a) are also shown. Black is an integration of the intensity  in the central spot of the beam profile (i.e. from radius = 0 $\mu$m to the first radial zero of the intensity), and red is the on-axis lineout (radius = 0 $\mu$m), normalized to their max values. Also shown are two transverse beam profiles taken at z = 70 mm (b), and z = 190mm (c).  These positions are marked with dashed black lines in (a). The profile at z = 70 mm corresponds to the peak of the first oscillation in the BG focal region from the axicon tip where maximum XUV and ionization yield is seen. The profile at z = 190 mm shows the vanishing on-axis intensity of a BG beam.}
\end{figure}

To better understand the oscillations in ionization and XUV yield, we have performed simulations for the three-dimensional propagation of the BG beams using a quasi-discrete Hankel transform method \cite{Guizar-Sicairos2004}. The simulation uses parameters comparable to those used in the experiment. Figure 4 shows the simulated intensity profile as the beam propagates from right to left in the figure. The figure also shows the on-axis intensity as a function of propagation distance (red)  as well as the integrated intensity profile to the first radial zero of the BG transverse profile (black). As expected, both the integrated and peak intensities exhibit oscillations typical of Bessel-like beams \cite{Akturk2008} that closely follow the measured ionization oscillations, however, the maximum of both curves occur at different points. The integrated intensity is maximized at the start of the focal region (z $\approx$ 10 mm in Figure 4), closer to the axicon tip, and degrades rapidly moving away from the axicon. In contrast, the on-axis intensity is more constant across the focal region and is maximized towards the end of the region (z $\approx$ 110 mm in Figure 4), further from the axicon. Comparing experiment to simulation, the HHG and ionization yields presented in Figure 3 mirror the integrated intensity profile in Figure 4 rather than the on-axis (peak) intensity profile. Therefore, we argue that the maximum ionization yield occurs at the first intensity oscillation in the BG focus from the axicon due to  volume averaging effects between the gas jet and focal profile. The XUV yield follows this as HHG is macroscopic, coherent process which scales with the number of emitters squared \cite{Shiner2009,Shiner}.
\\
With the simulation presented in Figure 4 some light is also shed on the ionization data previously touched on. Horizontal red dashed lines in Figure 3 mark the ionization level equivalent to the value seen at the position along the focal region of maximum XUV yield and black circles mark other positions where there is similar ionization yield. Though there are multiple positions where this same ionization yield occurs, they vary drastically in XUV yield, particularly in decreasing at positions further from the axicon tip. To understand this observation we need to remind the reader the experimental limitation of our ionization measurements. By using a channeltron we are not able to measure the ionization fraction. Instead we can only conclude that to achieve a similar current in the channeltron output but from a smaller interaction region, a larger fraction of ions need to be generated. Figure 4 shows that the size of the central spot in the BG focal region shrinks moving away from the axicon while the peak intensity remains largely constant with each intensity oscillation. This indicates that the ionization fraction increases with each oscillation in intensity as the peak intensity remains similar while the spot size shrinks for a constant gas target. Thus, HHG phase matching is degraded with each oscillation due to the presence of a higher concentration of ions. Strong field ionization and macroscopic harmonic simulations are needed to ratify our experimental findings.

\section{Concluding remarks}

We measured higher-order harmonics and ionization yields in unison from Bessel-like beams in a thin Ar gas jet, which is scanned along the propagation axis. The BG beams are created with a lens-axicon optical system that provides flexibility of the spot size, working distance and focal range when compared with Gaussian optics. The ionization follows the intensity oscillations in BG beams created by non-perfect axicons and subsequently there are multiple regions along these oscillations of favorable conditions for HHG. Harmonics and ionization are both maximum around the first oscillatory point where there is a lower on-axis peak intensity. We attribute the larger yields to a larger volume integral at this spatial point. This surprising result is complimented by numerical simulations of BG beams produced by a comparable lens and axicon system. The numerical simulations show that the integrated field strength over the center part of the spatial distribution is indeed larger for the initial oscillation compared to the spatial oscillation at the highest peak intensity.
\\
Finally, despite the driving field having a donut-like far field beam profile, the HHG beam is well collimated in the far field. Thus, generation of attosecond pulses with BG beams presents a novel spatially tunable, self-filtering approach for the strong field needed for the HHG. Besides removing the need of spatial filters, the presence of multiple regions for XUV generation also opens the door for multiple interacting regions as an approach for the generation of attosecond pulses.

\section{Acknowledgements}

This work was funded by the Air Force Office of Scientific Research, Grant No. FA9550-17-1-0369. T.S. was partially supported by the Chemical Sciences, Geosciences, and Biosciences Division, Office of Basic Energy Sciences, Office of Science, U.S. Department of Energy (DOE) Grant No. DE-SC0019098.

\section{Disclosures}
The authors declare no conflicts of interest.


\bibliography{library}

\begin{thebibliography}{10}
\newcommand{\enquote}[1]{``#1''}

\bibitem{McPherson1987}
A.~McPherson, G.~Gibson, H.~Jara, U.~Johann, T.~S. Luk, I.~A. McIntyre,
  K.~Boyer, and C.~K. Rhodes, \enquote{{Studies of multiphoton production of
  vacuum-ultraviolet radiation in the rare gases},} {\protect\JournalTitle{J.
  Opt. Soc. Am. B}} \textbf{4}, 595 (1987).

\bibitem{LHuillier1993}
A.~L'Huillier and P.~Balcou, \enquote{{High-order harmonic generation in rare
  gases with a 1-ps 1053-nm laser},} {\protect\JournalTitle{Physical Review
  Letters}} \textbf{70}, 774--777 (1993).

\bibitem{Krause1992}
J.~L. Krause, K.~J. Schafer, and K.~C. Kulander, \enquote{{High-order harmonic
  generation from atoms and ions in the high intensity regime},}
  {\protect\JournalTitle{Physical Review Letters}} \textbf{68}, 3535--3538
  (1992).

\bibitem{Corkum1993}
P.~B. Corkum, \enquote{{Plasma perspective on strong field multiphoton
  ionization},} {\protect\JournalTitle{Phys. Rev. Lett.}} \textbf{71},
  1994--1997 (1993).

\bibitem{Corkum2007}
P.~Corkum and F.~Krausz, \enquote{Attosecond science,}
  {\protect\JournalTitle{Nature physics}} \textbf{3}, 381--387 (2007).

\bibitem{Krausz2009a}
F.~Krausz and M.~Ivanov, \enquote{Attosecond physics,}
  {\protect\JournalTitle{Rev. Mod. Phys.}} \textbf{81}, 163--234 (2009).

\bibitem{Gaumnitz2017}
T.~Gaumnitz, A.~Jain, Y.~Pertot, M.~Huppert, I.~Jordan, F.~Ardana-Lamas, and
  H.~J. W{\"{o}}rner, \enquote{{Streaking of 43-attosecond soft-X-ray pulses
  generated by a passively CEP-stable mid-infrared driver},}
  {\protect\JournalTitle{Opt. Express}} \textbf{25}, 27506 (2017).

\bibitem{Popmintchev2010}
T.~Popmintchev, M.-C. Chen, P.~Arpin, M.~M. Murnane, and H.~C. Kapteyn,
  \enquote{{The attosecond nonlinear optics of bright coherent X-ray
  generation},} {\protect\JournalTitle{Nat. Photonics}} \textbf{4}, 822--832
  (2010).

\bibitem{Goulielmakis2010}
E.~Goulielmakis, Z.~H. Loh, A.~Wirth, R.~Santra, N.~Rohringer, V.~S. Yakovlev,
  S.~Zherebtsov, T.~Pfeifer, A.~M. Azzeer, M.~F. Kling, S.~R. Leone, and
  F.~Krausz, \enquote{{Real-time observation of valence electron motion},}
  {\protect\JournalTitle{Nature}} \textbf{466}, 739--743 (2010).

\bibitem{Uiberacker2007}
M.~Uiberacker, T.~Uphues, M.~Schultze, A.~J. Verhoef, V.~Yakovlev, M.~F. Kling,
  J.~Rauschenberger, N.~M. Kabachnik, H.~Schr{\"{o}}der, M.~Lezius, K.~L.
  Kompa, H.~G. Muller, M.~J. Vrakking, S.~Hendel, U.~Kleineberg, U.~Heinzmann,
  M.~Drescher, and F.~Krausz, \enquote{{Attosecond real-time observation of
  electron tunnelling in atoms},} {\protect\JournalTitle{Nature}} \textbf{446},
  627--632 (2007).

\bibitem{Schultze2010}
M.~Schultze, M.~Fie{\ss}, N.~Karpowicz, J.~Gagnon, M.~Korbman, M.~Hofstetter,
  S.~Neppl, A.~L. Cavalieri, Y.~Komninos, T.~Mercouris, C.~A. Nicolaides,
  R.~Pazourek, S.~Nagele, J.~Feist, J.~Burgd{\"{o}}rfer, A.~M. Azzeer,
  R.~Ernstorfer, R.~Kienberger, U.~Kleineberg, E.~Goulielmakis, F.~Krausz, and
  V.~S. Yakovlev, \enquote{{Delay in photoemission},}
  {\protect\JournalTitle{Science}} \textbf{328}, 1658--1662 (2010).

\bibitem{Attar2017}
A.~R. Attar, A.~Bhattacherjee, C.~D. Pemmaraju, K.~Schnorr, K.~D. Closser,
  D.~Prendergast, and S.~R. Leone, \enquote{{Femtosecond x-ray spectroscopy of
  an electrocyclic ring-opening reaction},} {\protect\JournalTitle{Science}}
  \textbf{356}, 54--59 (2017).

\bibitem{Geneaux2019}
R.~Geneaux, H.~J. Marroux, A.~Guggenmos, D.~M. Neumark, and S.~R. Leone,
  \enquote{Transient absorption spectroscopy using high harmonic generation: a
  review of ultrafast x-ray dynamics in molecules and solids,}
  {\protect\JournalTitle{Philosophical Transactions of the Royal Society A}}
  \textbf{377}, 20170463 (2019).

\bibitem{Kraus2018}
P.~M. Kraus, M.~Z{\"{u}}rch, S.~K. Cushing, D.~M. Neumark, and S.~R. Leone,
  \enquote{{The ultrafast X-ray spectroscopic revolution in chemical
  dynamics},} {\protect\JournalTitle{Nat. Rev. Chem.}} \textbf{2}, 82--94
  (2018).

\bibitem{Cavalieri2007}
A.~L. Cavalieri, N.~M{\"{u}}ller, T.~Uphues, V.~S. Yakovlev, A.~Baltu{\v{s}}ka,
  B.~Horvath, B.~Schmidt, L.~Bl{\"{u}}mel, R.~Holzwarth, S.~Hendel,
  M.~Drescher, U.~Kleineberg, P.~M. Echenique, R.~Kienberger, F.~Krausz, and
  U.~Heinzmann, \enquote{{Attosecond spectroscopy in condensed matter},}
  {\protect\JournalTitle{Nature}} \textbf{449}, 1029--1032 (2007).

\bibitem{Silva2015}
F.~Silva, S.~M. Teichmann, S.~L. Cousin, M.~Hemmer, and J.~Biegert,
  \enquote{{Spatiotemporal isolation of attosecond soft X-ray pulses in the
  water window},} {\protect\JournalTitle{Nat. Commun.}} \textbf{6}, 6611
  (2015).

\bibitem{Young2018}
L.~Young, K.~Ueda, M.~G{\"u}hr, P.~H. Bucksbaum, M.~Simon, S.~Mukamel,
  N.~Rohringer, K.~C. Prince, C.~Masciovecchio, M.~Meyer \emph{et~al.},
  \enquote{Roadmap of ultrafast x-ray atomic and molecular physics,}
  {\protect\JournalTitle{Journal of Physics B: Atomic, Molecular and Optical
  Physics}} \textbf{51}, 032003 (2018).

\bibitem{Vampa2015}
G.~Vampa, T.~J. Hammond, N.~Thir{\'{e}}, B.~E. Schmidt, F.~L{\'{e}}gar{\'{e}},
  C.~R. McDonald, T.~Brabec, and P.~B. Corkum, \enquote{{Linking high harmonics
  from gases and solids},} {\protect\JournalTitle{Nature}} \textbf{522},
  462--464 (2015).

\bibitem{McLeod1954}
J.~H. McLeod, \enquote{{The Axicon: A New Type of Optical Element},}
  {\protect\JournalTitle{J. Opt. Soc. Am.}} \textbf{44}, 592 (1954).

\bibitem{McLeod1960}
J.~H. McLeod, \enquote{{Axicons and Their Uses},} {\protect\JournalTitle{J.
  Opt. Soc. Am.}} \textbf{50}, 166 (1960).

\bibitem{Durnin1987}
J.~Durnin, J.~Miceli, and J.~H. Eberly, \enquote{{Diffraction-free beams},}
  {\protect\JournalTitle{Phys. Rev. Lett.}} \textbf{58}, 1499--1501 (1987).

\bibitem{Durnin1988}
J.~Durnin, J.~H. Eberly, and J.~J. Miceli, \enquote{{Comparison of Bessel and
  Gaussian beams},} {\protect\JournalTitle{Opt. Lett.}} \textbf{13}, 79 (1988).

\bibitem{Summers2017}
A.~M. Summers, X.~Yu, X.~Wang, M.~Raoul, J.~Nelson, D.~Todd, S.~Zigo, S.~Lei,
  and C.~A. Trallero-Herrero, \enquote{{Spatial characterization of Bessel-like
  beams for strong-field physics},} {\protect\JournalTitle{Opt. Express}}
  \textbf{25}, 1646 (2017).

\bibitem{akturk2012tailored}
S.~Akturk, \enquote{Tailored-beam ultrashort laser pulses,}
  {\protect\JournalTitle{Quantum. Phys. Lett.}} \textbf{1}, 97--112 (2012).

\bibitem{Mcgloin2005}
D.~Mcgloin and K.~Dholakia, \enquote{{Bessel beams: Diffraction in a new
  light},} {\protect\JournalTitle{Contemp. Phys.}} \textbf{46}, 15--28 (2005).

\bibitem{Jarutis2000}
V.~Jarutis, R.~Pa{\v{s}}kauskas, and A.~Stabinis, \enquote{{Focusing of
  Laguerre-Gaussian beams by axicon},} {\protect\JournalTitle{Opt. Commun.}}
  \textbf{184}, 105--112 (2000).

\bibitem{Grunwald2008}
R.~Grunwald, M.~Bock, V.~Kebbel, S.~Huferath, U.~Neumann, G.~Steinmeyer,
  G.~Stibenz, J.-L. N{\'{e}}ron, and M.~Pich{\'{e}},
  \enquote{{Ultrashort-pulsed truncated polychromatic Bessel-Gauss beams},}
  {\protect\JournalTitle{Opt. Express}} \textbf{16}, 1077 (2008).

\bibitem{lin_le_jin_wei_2018}
C.~D. Lin, A.-T. Le, C.~Jin, and H.~Wei, \emph{Attosecond and Strong-Field
  Physics: Principles and Applications} (Cambridge University Press, 2018).

\bibitem{Luo2015}
J.~Luo, Q.~Cheng, and D.~Xu, \enquote{{Generation of intense isolated
  attosecond pulses with high spatiotemporal quality by two-color
  polarization-gating Bessel-Gauss beams},} {\protect\JournalTitle{Opt.
  Commun.}} \textbf{339}, 247--253 (2015).

\bibitem{Wang2012}
Z.~Wang, W.~Hong, Q.~Zhang, S.~Wang, and P.~Lu, \enquote{{Efficient generation
  of isolated attosecond pulses with high beam quality by two-color
  Bessel–Gauss beams},} {\protect\JournalTitle{Opt. Lett.}} \textbf{37}, 238
  (2012).

\bibitem{Li2012}
Y.~Li, Q.~Zhang, W.~Hong, S.~Wang, Z.~Wang, and P.~Lu, \enquote{{Efficient
  generation of high beam-quality attosecond pulse with polarization-gating
  Bessel-Gauss beam from highly-ionized media},} {\protect\JournalTitle{Opt.
  Express}} \textbf{20}, 15427 (2012).

\bibitem{VanDao2009}
L.~{Van Dao}, K.~B. Dinh, and P.~Hannaford, \enquote{{Generation of extreme
  ultraviolet radiation with a Bessel-Gaussian beam},}
  {\protect\JournalTitle{Appl. Phys. Lett.}} \textbf{95} (2009).

\bibitem{Averchi2008}
A.~Averchi, D.~Faccio, R.~Berlasso, M.~Kolesik, J.~V. Moloney, A.~Couairon, and
  P.~{Di Trapani}, \enquote{{Phase matching with pulsed Bessel beams for
  high-order harmonic generation},} {\protect\JournalTitle{Phys. Rev. A}}
  \textbf{77}, 021802 (2008).

\bibitem{Klas2018}
R.~Klas, A.~Kirsche, M.~Tschernajew, J.~Rothhardt, and J.~Limpert,
  \enquote{{Annular beam driven high harmonic generation for high flux coherent
  XUV and soft X-ray radiation},} {\protect\JournalTitle{Opt. Express}}
  \textbf{26}, 19318 (2018).

\bibitem{Akturk2008}
S.~Akturk, B.~Zhou, B.~Pasquiou, M.~Franco, and A.~Mysyrowicz,
  \enquote{{Intensity distribution around the focal regions of real axicons},}
  {\protect\JournalTitle{Opt. Commun.}} \textbf{281}, 4240--4244 (2008).

\bibitem{Shiner}
A.~D. Shiner, C.~Trallero-Herrero, N.~Kajumba, B.~E. Schmidt, J.~B. Bertrand,
  K.~T. Kim, H.-C. Bandulet, D.~Comtois, J.-C. Kieffer, D.~M. Rayner, P.~B.
  Corkum, F.~L{\'{e}}gar{\'{e}}, and D.~M. Villeneuve, \enquote{{High harmonic
  cutoff energy scaling and laser intensity measurement with a 1.8 $\mu$m laser
  source},} {\protect\JournalTitle{Journal of Modern Optics}} .

\bibitem{Shiner2009}
A.~D. Shiner, C.~Trallero-Herrero, N.~Kajumba, H.~C. Bandulet, D.~Comtois,
  F.~L{\'{e}}gar{\'{e}}, M.~Gigu{\`{e}}re, J.~C. Kieffer, P.~B. Corkum, and
  D.~M. Villeneuve, \enquote{{Wavelength scaling of high harmonic generation
  efficiency},} {\protect\JournalTitle{Phys. Rev. Lett.}} \textbf{103} (2009).

\bibitem{Henke1993}
B.~L. Henke, E.~M. Gullikson, and J.~C. Davis, \enquote{{X-ray interactions:
  Photoabsorption, scattering, transmission, and reflection at E = 50-30, 000
  eV, Z = 1-92},} {\protect\JournalTitle{At. Data Nucl. Data Tables}}
  \textbf{54}, 181--342 (1993).

\bibitem{tross2017}
J.~Tross, G.~Kolliopoulos, and C.~A. Trallero-Herrero, \enquote{A
  self-referencing attosecond interferometer,} {\protect\JournalTitle{APS}}
  \textbf{2017}, Q1--139 (2017).

\bibitem{Trob2019}
J.~Tro{\ss} and C.~A. Trallero-Herrero, \enquote{{High harmonic generation
  spectroscopy via orbital angular momentum},} {\protect\JournalTitle{J. Chem.
  Phys.}} \textbf{151}, 84308 (2019).

\bibitem{Guizar-Sicairos2004}
M.~Guizar-Sicairos and J.~C. Guti{\'{e}}rrez-Vega, \enquote{{Computation of
  quasi-discrete Hankel transforms of integer order for propagating optical
  wave fields},} {\protect\JournalTitle{J. Opt. Soc. Am. A}} \textbf{21}, 53
  (2004).

\end{thebibliography}






\end{document}